# Nonlocal effective medium approximation for metallic nanorod metamaterials

Tao Geng[1,2], Songlin Zhuang[2], Jie Gao[1], Xiaodong Yang[1*]

[1] Department of Mechanical and Aerospace Engineering, Missouri University of Science and Technology, Rolla, Missouri 65409, USA

[2] Engineering Research Center of Optical Instrument and System, Ministry of Education, Shanghai Key Lab of Modern Optical System, University of Shanghai for Science and Technology, Shanghai 200093, China

*E-mail: yangxia@mst.edu

PACS number(s): 78.20.Ci, 42.70.−a, 78.20.Bh

We present an analytical nonlocal effective medium approximation to describe the optical nonlocal effects in metallic nanorod metamaterials based on Mie scattering theory. It is shown that the developed nonlocal effective medium theory can predict a coexistence state of two modes around the epsilon-near-zero region, where strong optical nonlocal effects lead to the behavior of both positive refraction and negative refraction in the nanorod metamaterials. Outside of the coexistence region, only one mode can be excited and its behavior can be well described using the local effective medium theory.

**I. INTRODUCTION**

Metallic nanorod metamaterials, which are composed of an array of aligned metallic nanorods embedded in a dielectric matrix, are one kind of uniaxial media with optical axis parallel to the direction of the nanorods, as shown in Fig.1 (a). Metallic nanorod metamaterials have recently been used in many applications due to their extraordinary optical properties, including negative refraction [1], subwavelength imaging [2, 3], biosensing [4], strong Purcell effect [5, 6], and Cherenkov emission [7].

In particular, metallic nanorod metamaterials exhibit strong optical nonlocality near the epsilon-near-zero (ENZ) regime, leading to the excitation of additional optical waves. Optical nonlocality and additional waves in ENZ metamaterials have been studied extensively in the literature [8-12]. The effective nonlocal permittivity along the nanorod direction $\varepsilon_z$ has been obtained by fitting to the three-dimensional finite element method (FEM) solutions [8]. Furthermore, a first-principle theoretical method has been proposed to achieve an explicit expression of $\varepsilon_z$ [9, 10], but the method is based on the assumption of ultrathin high-conductivity wires and it may not be applied at visible and near-IR frequencies. Then, a quasistatic analytical model of wire media has been developed in Ref. [12]. However, in contrast with the method proposed in Ref. [9], the model in Ref. [12] does not provide high accuracy for the thin wires system. In addition, another theoretical method utilizing the cylindrical surface plasmon modes is presented in Ref. [11] by providing an explicit relation between $k_z$ and $k_x$ to describe the nonlocal effects, but other important parameters have to be solved from a series of equations.

In this work, we develop an analytical nonlocal effective medium theory (EMT) to describe the effective permittivity of the metallic nanorod metamaterials based on Mie scattering theory and also verify the theory with numerical simulation (COMSOL Multiphysics). Previous experimental work [8] has been used to analyze the proposed nonlocal EMT in detail. The microscopic field distributions obtained from the numerical simulation have been used to confirm the analytical results.

## II. DEVELOPMENT OF NONLOCAL EFFECTIVE MEDIUM THEORY

Here the material permeabilities are assumed to be unity in optical frequencies and the condition of $\lambda \gg a$ is satisfied. It is assumed that the incident plane is the xz plane as shown in Fig. 1(a). For TE polarization (electric field is along the *y* direction), the metamaterial can be treated as an isotropic homogeneous medium due to the fact of $\varepsilon_x = \varepsilon_y$, so that the dispersion relation is

$$k_z^2 + k_x^2 = \varepsilon_{x(y)} k^2 \qquad (1)$$

where *k* is the wave number in vacuum. The effective permittivity $\varepsilon_{x(y)}$ is related to the plasmonic oscillations perpendicular to the nanorod axes (transverse modes) [13]. In this work, the interested region with strong nonlocal effective permittivity $\varepsilon_z$ is located outside of the transverse mode region, so that $\varepsilon_{x(y)}$ can be well described by the Maxwell Garnett (MG) effective medium theory (EMT) in long wavelength limit without considering the nonlocal properties [8, 10, 11]. For TM polarization (magnetic field is along the *y* direction), the dispersion relation is,

$$\frac{k_x^2}{\varepsilon_z} + \frac{k_z^2}{\varepsilon_{x(y)}} = k^2 \qquad (2)$$

In this case, $k_z$ cannot be obtained explicitly from Eq. (2), because the nonlocal effective permittivity $\varepsilon_z$ is a function of $k_z$.

The metallic nanorods with permittivity $\varepsilon_m$ and radius $r_m$ are arranged in a periodic square lattice or hexagonal lattice, embedded in a background dielectric medium with permittivity $\varepsilon_h$. It should be pointed out that the method is still applicable even when the position disorder is introduced in the periodic lattice. For simplicity, a square lattice has been taken into account in the following analysis, as shown in Fig.1 (a). The light propagation in a single nanorod unit cell can be considered as the scattering in an effective medium with permittivity $\varepsilon_{eff}$ consisting of the metallic nanorod as the inner core and a coated cylinder of radius $r_h$ with the background medium. The unit cell is then approximated by a circular disk having the same area as the coated cylinder ($\pi r_h^2 = a^2$), as shown in Fig.1 (b). Effective permittivity $\varepsilon_{eff}$ can then be determined by the condition that the total scattering vanishes in the limit of $k_{eff} r_h \ll 1$ where $k_{eff}$ is the wave number in the effective medium. This method is so called the coherent-potential approximation (CPA) [14], which has been used to retrieve the effective parameters for the two-dimensional photonic crystals [15-17] and acoustic crystals [18]. In this work, the CPA method has been extended to discuss the oblique incidence of electromagnetic wave onto the metallic nanorod metamaterial surface.

To study the nonlocal nature of TM mode, it has been considered that a plane wave with magnetic field along the y direction illuminates the metallic nanorod metamaterials. The scattered electromagnetic wave in Fig. 1(b) can be expanded by vector cylindrical wave functions [19] when $L \gg r_h$. The electric field along the z direction in the metallic rod can be written as

$$E_z(k_{mx}, r) = \frac{k_{mx}^2}{\sqrt{\varepsilon_m} k} \sum_n a_n^m J_n(k_{mx} r) e^{-ik_z z + in\phi} \qquad (3)$$

In the coated cylinder layer, the corresponding electric field component is

$$E_z(k_{hx}, r) = \frac{k_{hx}^2}{\sqrt{\varepsilon_h} k} \sum_n \left[ a_n^h J_n(k_{kx} r) + b_n^h H_n^{(1)}(k_{hk} r) \right] e^{-ik_z z + in\phi} \qquad (4)$$

Finally, the electric field along the z direction in the effective medium has the expression of

$$E_z(k_{ex}, r) = \frac{k_{ex}^2}{\sqrt{\varepsilon_z} k} \sum_n \left[ a_n^{eff} J_n(k_{ex} r) + b_n^{eff} H_m^{(1)}(k_{ex} r) \right] e^{-ik_z z + in\phi} \quad (5)$$

Here $J_n(x)$ and $H_n^{(1)}(x)$ are the Bessel function and Hankel function of the first kind with $k_{\{m,h,e\}x}^2 + k_z^2 = \varepsilon_{\{m,h,z\}} k^2$. It is noted that $k_{ex}$ is not $k_x$ in Eq. (2) but is a transverse wave vector of the mode in the microscopic theory. Then the corresponding magnetic field can be obtained by using $\vec{H} = -\frac{i}{\omega \mu_0} \nabla \times \vec{E}$, where $\mu_0$ is the permeability in vacuum.

The total scattering cross section of the coated cylinder layer is given by [14, 19]

$$C_{sca} = \frac{4}{k_{eff}} \sum_n |D_n^{eff}|^2 \quad (6)$$

where $D_n^{eff} = b_n^{eff} / a_n^{eff}$ is the Mie scattering coefficients of the whole coated cylinder layer. Taking into account the condition that the total scattering vanishes in the effective medium, it has $D_n^{eff} = 0$, or, equivalently, $b_n^{eff} = 0$. Then by matching the boundary conditions at $r = r_m$ and $r = r_h$, it can be obtained that

$$\frac{b_n^h}{a_n^h} = \frac{-J_n'(k_{hx} r_m) J_n(k_{mx} r_m) k_{hx} + J_n'(k_{mx} r_m) J_n(k_{hx} r_m) k_{mx}}{-J_n'(k_{mx} r_m) H_n^{(1)}(k_{hx} r_m) k_{mx} + H_n^{(1)'}(k_{hx} r_m) J_n(k_{mx} r_m) k_{hx}}$$
$$= \frac{-J_n'(k_{hx} r_h) J_n(k_{ex} r_h) k_{hx} + J_n'(k_{ex} r_h) J_n(k_{hx} r_h) k_{ex}}{-J_n'(k_{ex} r_h) H_n^{(1)}(k_{hx} r_h) k_{ex} + H_n^{(1)'}(k_{hx} r_h) J_n(k_{ex} r_h) k_{hx}} \quad (7)$$

Considering the limit of $k_{eff} r_h \ll 1$, for TM polarization, the effective permittivity is dominated by the $n = 0$ term [14]. Thus from Eq. (7), it has the form of

$$\frac{-J_0'(k_{hx} r_m) J_0(k_{mx} r_m) k_{hx} + J_0'(k_{mx} r_m) J_0(k_{hx} r_m) k_{mx}}{-J_0'(k_{mx} r_m) H_0^{(1)}(k_{hx} r_m) k_{mx} + H_0^{(1)'}(k_{hx} r_m) J_0(k_{mx} r_m) k_{hx}} = \frac{-J_0'(k_{hx} r_h) J_0(k_{ex} r_h) k_{hx} + J_0'(k_{ex} r_h) J_0(k_{hx} r_h) k_{ex}}{-J_0'(k_{ex} r_h) H_0^{(1)}(k_{hx} r_h) k_{ex} + H_0^{(1)'}(k_{hx} r_h) J_0(k_{ex} r_h) k_{hx}}$$
(8)

In order to obtain an explicit function of the nonlocal effective permittivity $\varepsilon_z$, an appropriate approximation should be involved. Considering the quasistatic regime ($a/\lambda \ll 1$), it has $k_{\{m,h,e\}x} r_{\{m,h\}} \ll 1$. In this case, the approximate expressions of $J_0(x)$, $J_0'(x)$, $H_0^{(1)}(x)$ and $H_0^{(1)'}(x)$ are $J_0(x) \approx 1$, $J_0'(x) \approx -x/2$, $H_0^{(1)}(x) \approx 1 + i\left( \frac{2 C_{Euler}}{\pi} + \frac{2 \ln(x/2)}{\pi} \right)$ and

$$H_0^{(1)\prime}(x) \approx -\frac{x}{2} + i\left(\frac{2}{\pi x} - \frac{3C_{Euler}x}{2\pi} - \frac{x\ln(x/2)}{\pi}\right) \quad \text{with} \quad x = k_{\{m,h,e\}x}r_{\{m,h\}}, \text{ respectively. Eq. (8) can then be}$$

simplified into

$$\varepsilon_z = \frac{\varepsilon_z^{mg} + q\varepsilon_m k_z^2}{1 + qk_z^2} \quad (9)$$

where the parameter $q = C_{Euler} p r_h^2 / 4$, $C_{Euler}$ is the Euler constant, $p$ is the surface concentration of metallic nanorods, and $\varepsilon_z^{mg} = (1-p)\varepsilon_h + p\varepsilon_m$ based on EMT. Substitute Eq. (9) into Eq. (2) and use the effective permittivity component derived from MG EMT $\varepsilon_x = \varepsilon_x^{mg} = \varepsilon_h \frac{(1+p)\varepsilon_m + (1-p)\varepsilon_h}{(1+p)\varepsilon_h + (1-p)\varepsilon_m}$, the dispersion relation can then be rewritten as

$$k_z^2 = -\frac{\varepsilon_z^{mg} + q\varepsilon_x^{mg}(k_x^2 - \varepsilon_m k^2) \pm \sqrt{-4q\varepsilon_m \varepsilon_x^{mg}(k_x^2 - \varepsilon_z^{mg}k^2) + \left(\varepsilon_z^{mg} + q\varepsilon_x^{mg}(k_x^2 - \varepsilon_m k^2)\right)^2}}{2q\varepsilon_m} \quad (10)$$

It is shown that there are two different solutions for $k_z$ in Eq. (10). In other words, there are two TM modes propagating inside the metallic nanorod metamaterials. It should be noted that Eq. (9) and Eq. (10) are valid when $k_{\{m,h,e\}x}r_{\{m,h\}} \ll 1$. As a result, the proposed theoretical model may not be valid at the IR regime and at lower frequencies, due to the fact that the absolute value of the permittivity of metal increases rapidly as frequency decreases.

### III. ANALYSIS OF NONLOCAL EFFECTS IN METALLIC NANOROD METAMATERIALS

The previous experimental work on metallic nanorod metamaterials [8] will be analyzed with the present nonlocal dispersion relation in Eq. (10). Based on Ref. [8], the parameters of the gold nanorod metamaterials are $r_m = 12.5nm$, $a = 60nm$, $L = 300nm$, $\varepsilon_h = 2.74$ (Al$_2$O$_3$) and $p = 2\pi r_m^2 / (\sqrt{3}a^2)$ (hexagonal lattice). The permittivity of gold nanorod is described by $\varepsilon_m = \varepsilon_{bulk} + \frac{i\omega_p^2 \tau(R_b - R)}{\omega(\omega\tau + i)(\omega\tau R + iR_b)}$ [8], where $\varepsilon_{bulk}$ is the permittivity of bulk gold [20], $R_b \approx 35.7nm$ is the mean free path of the electrons in bulk gold, $\omega_p \approx 13.7 \times 10^{15} Hz$ is the plasma frequency, $\tau = 2.53 \times 10^{-14} s$ is the relaxation time for the free electrons in gold, and $R$ is the effective mean free path equal to $5nm$ and $10nm$ for the unannealed and annealed samples, respectively. According to Ref. [8], the real part of gold permittivity changes little as $R$ increasing, but the imaginary part decrease a lot. Then, the dispersion relation of the annealed sample with $60^0$ incident angle can be calculated with Eq. (10), as shown in Fig.2. It can be seen that the main mode shifts from $k_z^{(1)}$ to $k_z^{(2)}$ as the wavelength increases. The similar results based on FEM simulations have been given in Fig. 2 (b) of Ref. [8].

For TM polarization, the incident magnetic field is along the y direction and can be written as

$$H_y = H_0 e^{ik_x x} \begin{cases} e^{ik_{0z}z} + \rho e^{-ik_{0z}z} & z < 0 \\ A_1 e^{ik_z^{(1)}z} + A_2 e^{-ik_z^{(1)}z} + B_1 e^{ik_z^{(2)}z} + B_2 e^{-ik_z^{(2)}z} & 0 < z < L \\ t e^{ik_{0z}(z-L)} & z > L \end{cases} \quad (11)$$

where $H_0$ is amplitude, $\rho$ is the reflection coefficient, $t$ is the transmission coefficient, $k_{0z}$ is the z component of the wave vector in vacuum, and $A_1$, $A_2$, $B_1$, and $B_2$ are the amplitudes of the TM modes in the metamaterial slab. Because of the existing two TM modes in such artificial material slab, an additional boundary condition (ABC) is needed to remove the additional degree of freedom [21, 22]. Previous studies [10, 23] have shown that at the interface of the metallic nanorod metamtaerial and an insulator, the ABC can be described as continuous $\varepsilon_h E_n$ at the interface, where $E_n$ is the normal component of the average electric field.

In Fig.3, the extinction spectra $-\ln(T)$ from the metallic nanorod metamaterials are calculated by incorporating the nonlocal EMT (Eq. (10)) with the ABC, where $T = |t|^2$ is the transmittivity using the transfer matrix method. These results are in agreement with the experiments shown in Ref. [8] (see Fig. 2 (c) and 2 (d) in Ref. [8]) and the FEM simulations. The results of MG EMT are also shown in Fig. 2 for comparison. There are two resonance peaks near 530nm and 650 nm, which are called transverse (T) mode and longitudinal (L) mode [8], respectively. It is clear that the major difference between the results from nonlocal EMT and MG EMT is in the L mode region. When the wavelength is out of the L mode region, the results from the both methods are almost the same. The difference between nonlocal EMT and MG EMT results will become more significant after the annealing process or when the incident angle increases. It is indicated that the nonlocal effects in the L mode region depend on both the absorption loss and incidence angle. Since the L mode is associated with a plasmonic excitation polarized along the nanorod long axes [24], it is excited more effectively at larger incident angles and with lower loss.

In order to understand the nonlocal optics in the metallic nanorod metamaterials, a lossless case will be discussed first. Fig. 4 shows the calculated isofrequency contour at different wavelengths using Eq. (10). At 400nm, $k_z^{(2)}$ is much larger than $k$, which means that mode 2 is hard to be excited by an incoming electromagnetic wave. $k_z^{(1)}$ is overlapped with $k_z^{mg}$ and it describes a normal positive refraction behavior due to the elliptic dispersion. Thereby, only mode 1 can be excited and it can be described well by MG EMT in this circumstance. The isofrequency contour of $k_z^{(2)}$ will become increasingly sharp as the wavelength increases. When the wavelength increases to 900nm, the dispersion curve of $k_z^{(1)}$ will be cutoff and only mode 2 will exist. $k_z^{(2)}$ is overlapped with $k_z^{mg}$ as $|k_x/k_0| \leq 1$ and it describes a negative refraction behavior due to the hyperbolic dispersion [25]. Thereby, only mode 2 can be excited and it can be described well by MG EMT in this circumstance. At 650nm, the values of $k_z^{(2)}$ is in the same order with $k_z^{(1)}$ as $|k_x/k_0| \leq 1$, which means that both mode 1 and mode 2 can be excited by an incoming electromagnetic wave. Thereby, there are two propagating modes in the metamaterial, where mode 1 and mode 2 describe the positive refraction and the negative refraction behavior, respectively. The nonlocal EMT must be considered in this circumstance.

Then, taking into account of the absorption loss of gold nanorods, we can get the same conclusion as the lossless case from the nonlocal effective permittivity $\varepsilon_z$. Two solutions of $\varepsilon_z$, $\varepsilon_z^{(1)}$ and $\varepsilon_z^{(2)}$, can be calculated by substituting Eq. (10) into Eq. (9). The real and imaginary parts of $\varepsilon_z^{(1)}$ and $\varepsilon_z^{(2)}$ with different incidence angles are shown in Fig. 5(a) and 5(b). In the short wavelength range, it has $\text{Re}(\varepsilon_z^{(1)}) > 0$ and $\varepsilon_z^{(2)} \approx 0$. It means that only mode 1 can be excited and the metamaterial slab exhibits positive refraction. In the long wavelength range, it has $\varepsilon_z^{(1)} \approx 0$ and $\text{Re}(\varepsilon_z^{(2)}) < 0$. This means that only mode 2 can be excited and the slab exhibits negative refraction. Around 650 nm, it has $\text{Re}(\varepsilon_z^{(1)}) > 0$ and $\text{Re}(\varepsilon_z^{(2)}) < 0$. This means that both mode 1 and mode 2 can be excited simultaneously and the metamaterial slab exhibits both positive refraction and negative refraction, leading to the strong optical nonlocal phenomenon observed in the experiments. The coexistence region of two modes becomes larger with the increased incidence angle due to the stronger plasmonic excitation polarized along the nanorod long axes. For local MG EMT, the dispersion curve of metallic nanorod metamaterials will change from elliptic to hyperbolic when the wavelength increases and crosses the ENZ position, as shown in Fig. 5(c). However, in fact the ENZ position is hard to be defined and it expands to a coexistence region of the two modes. The dispersion curve shows a "hybridized" feature in this coexistence region as shown in Fig. 4(b). Thus, the dispersion curve will change from elliptic to "hybridized" and then to hyperbolic when the wavelength increases as shown in Fig 5(d). In the elliptic or hyperbolic regime, only one mode can be excited and the dispersion can be described well by MG EMT. In the "hybridized" regime, nonlocal EMT must to be considered since two modes will coexist. These results can explain why the local MG EMT can be used well for silver nanorod metamaterials in the hyperbolic regime [26], even the loss of silver is smaller than gold.

A new parameter $\delta$ has been defined as $\delta = \frac{|B_1|^2 - |A_1|^2}{|B_1|^2 + |A_1|^2}$ in order to study the macroscopic behavior of the two modes propagating in the metamaterial slab, where $A_1$ and $B_1$ are defined in Eq. (11). The calculated $\delta$ of the annealed sample ($R$=10nm) for different incident angles using the transfer matrix method are shown in Fig. 6. As expected, in the elliptic regime, it has $\delta \approx -1$ due to $|B_1| \to 0$. And in the hyperbolic regime, it has $\delta \approx 1$ due to $|A_1| \to 0$. In the "hybridized" regime, both mode 1 and mode 2 can be excited simultaneously, leading to $1 < \delta < -1$. The region of $1 < \delta < -1$ will become larger as the incidence angle increases.

In fact, the nonlocal EMT represents one method of homogenization. In order to verify the validity of above discussion, we study the microscopic model of the metallic nanorod metamaterial slab using FEM simulations shown in Fig. 6. As expected, the *x*-component of time-averaged energy flow $S_x$ propagates along the +*x* direction in the elliptic regime (at 450nm) and along the -*x* direction in the hyperbolic regime (at 780nm). In the "hybridized" regime, it is clearly that $S_x$ is propagation along both the +*x* and -*x* directions simultaneously. And the dominant propagation direction of $S_x$ will change from +*x* to −*x* direction when the wavelength increases from 645nm to 675nm. At the same time, the *z*-component of time-averaged energy flow $S_z$ mainly propagates in the dielectric host and is always propagation along the +*z* direction at all wavelengths as shown in Fig. 7(b). The simulation results are in agreement with the macroscopic analysis using the nonlocal EMT presented in this work.

## IV. CONCLUSIONS

In conclusion, we develop a new type of nonlocal effective medium approximation approach to describe the optical nonlocal effects in metallic nanorod metamaterials. The derived nonlocal EMT can predict the optical nonlocal behavior in the previous experimental observation and also agree with the numerical simulations. It is demonstrated that the ENZ position in metallic nanorod metamaterials expands to a "hybridized" regime which cannot be described by local MG EMT. The coexistence of two modes in the "hybridized" regime leads to strong optical nonlocal effects, where the metamaterial exhibits both positive refraction and negative refraction behavior. Outside of the "hybridized" regime, only one mode can be excited which can be described well using local MG EMT.

**ACKNOWLEDGMENTS**

This work was supported by the University of Missouri Interdisciplinary Intercampus Research Program, the Ralph E. Powe Junior Faculty Enhancement Award, the National Science Foundation under grant CBET-1402743, the National Basic Research Program of China (2014CB339800), Shanghai Rising-Star Program (12QA1402300), and the National Natural Science Foundation of China (61008044). The authors acknowledge L. Sun for his helpful discussions about this work.

**FIGURES**

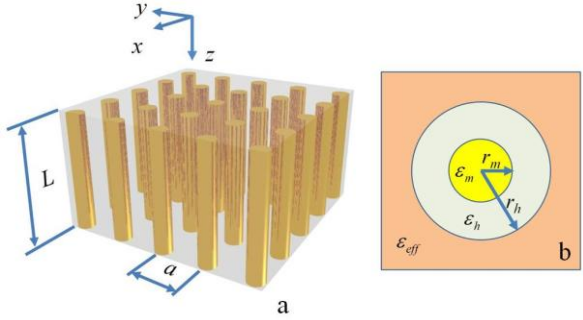

Fig. 1 (Color online) (a) Schematic geometry of metallic nanorod metamaterials in square lattice; (b) The equivalent unit cell of metamaterial in the nonlocal effective medium approximation.

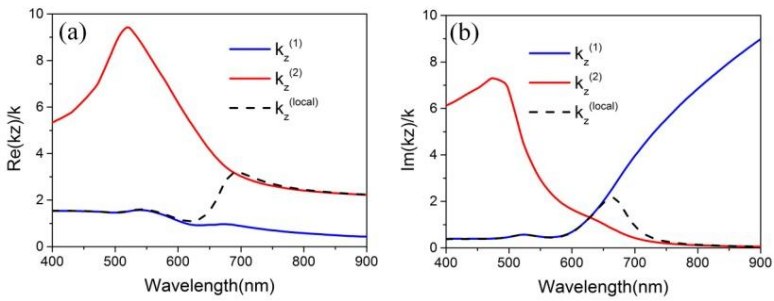

Fig.2 (Color online) Real part (a) and imaginary part (b) of the $k_z$ calculated from nonlocal EMT (Eq. (10), solid curves) and local EMT (Eq. (2), dashed curves) for the annealed sample ($R$=10nm) with $60^0$ incident angle in Ref. [8].

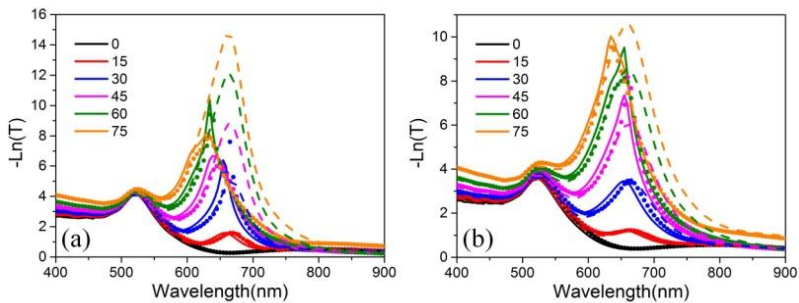

Fig.3 (Color online) Extinction spectra derived from nonlocal EMT (solid line), FEM-based simulations (dot) and local MG EMT (dashed line) for the annealed sample (a) and the unannealed sample (b).

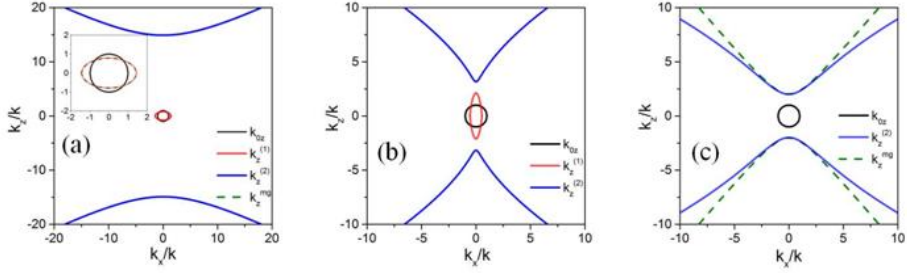

Fig. 4 (Color online) Calculated isofrequency contours using Eq. (10) for the lossless sample at the wavelengths of 400nm (a), 650nm (b) and 900nm (c). Dash lines in (a) and (c) show the calculated dispersion curves using local MG EMT. The insert figure in (a) shows the enlarged diagram of mode 1.

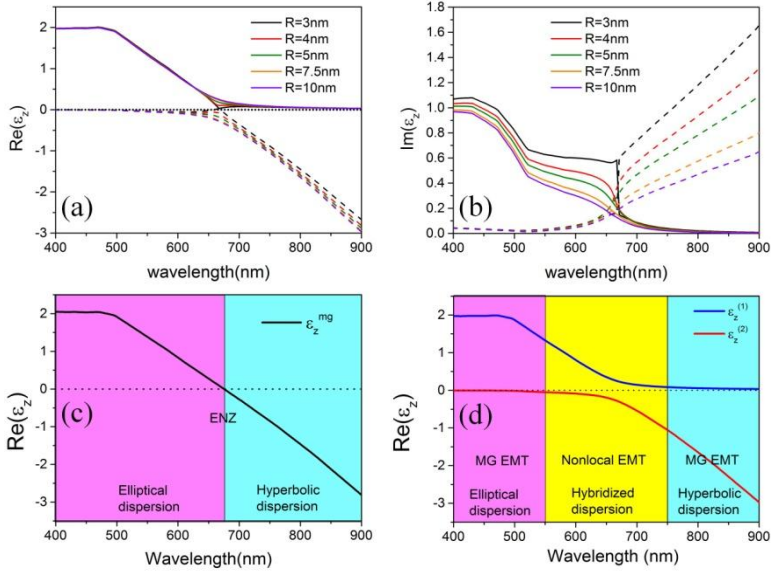

Fig. 5 (Color online) Real part (a) and imaginary part (b) of $\varepsilon_z^{(1)}$ and $\varepsilon_z^{(2)}$ for the annealed sample (R=10nm) at different incident angles. Solid curves and dash curves represent $\varepsilon_z^{(1)}$ and $\varepsilon_z^{(2)}$, respectively. (c, d) Schematic diagram of dispersion relation of metallic nanorod metamaterials for local MG EMT (c) and nonlocal EMT (d).

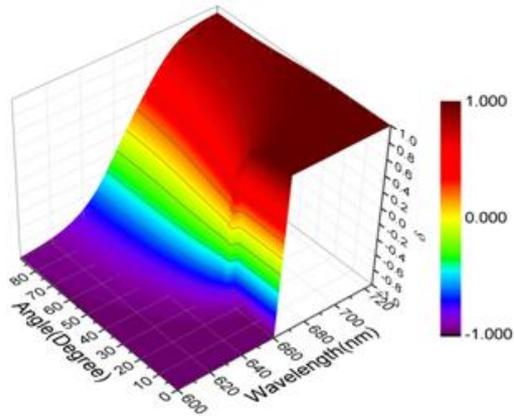

Fig. 6 (Color online) The calculated results of $\delta$ for the annealed sample (R=10nm) at difference incident angles.

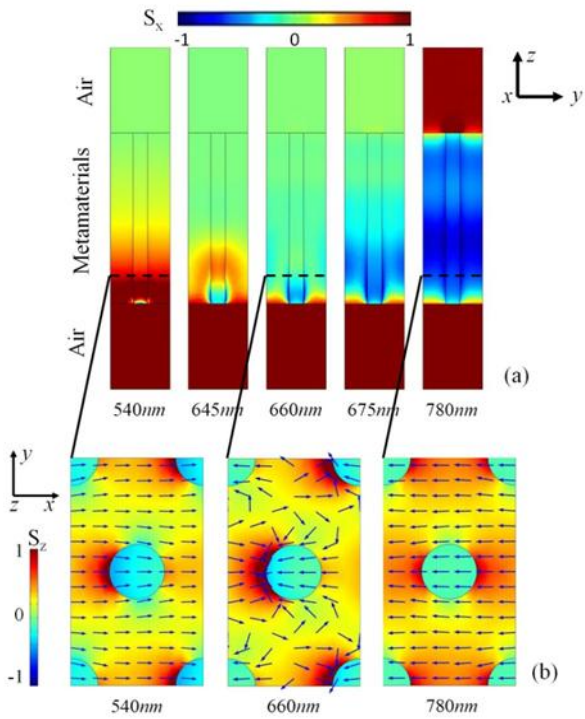

Fig. 7 (Color online) (a) The simulated x-components of time-averaged energy flow distributions $S_x$ for the annealed sample ($R$=10nm) with $60^0$ incident angle in the cross-section of yz plane at different wavelengths. The dash lines in (a) show the z positions of (b). (b) The simulated z-components of time-averaged energy flow distributions $S_z$ in the cross-section of xy plane at different wavelengths. The arrows in (b) show the directions of the energy flow within the xy plane.